\documentclass[preprint,showpacs,preprintnumbers,amsmath,amssymb]{revtex4}

\usepackage{graphicx}% Include figure files
\usepackage{dcolumn}% Align table columns on decimal point
\usepackage{bm}% bold math
\usepackage{latexsym}
\usepackage{amsfonts}
\usepackage{amssymb}
\usepackage{amsmath}
\begin{document}

%\preprint{KUNS-????}

\title{ Black String Perturbations in RS1 Model}

\author{Sugumi Kanno}
\email{sugumi@tap.scphys.kyoto-u.ac.jp}
\author{Jiro Soda}
\email{jiro@tap.scphys.kyoto-u.ac.jp}
\affiliation{
 Department of Physics,  Kyoto University, Kyoto 606-8501, Japan
}%

\date{\today}% It is always \today, today,

\begin{abstract}
 We present a general formalism for black string perturbations
  in Randall-Sundrum 1 model (RS1).
 First, we derive the master equation for the electric part of the 
 Weyl tensor $E_{\mu\nu}$.  Solving the master equation
  using the gradient expansion method, 
 we give the effective Teukolsky equation on the brane at low energy.
 It is useful to estimate gravitational waves emitted 
 by perturbed rotating black strings.
 We also argue the effect of the Gregory-Laflamme instability on
 the brane using our formalism.
\end{abstract}

\pacs{98.80.Cq, 98.80.Hw, 04.50.+h}% PACS, the Physics and Astronomy
\maketitle

\section{Introduction}

 The existence of extra-dimensions is the most exciting prediction
 of the superstring theory. The evidence of extra 
dimensions should be sought in the early stage of 
the universe or in the features of gravitational waves from the black hole. 
 We explore the latter possibility here.
 
In this paper, we concentrate on a two-brane model proposed 
by Randall and Sundrum as a simple  model~\cite{RS}. 
In this model, the black hole on the brane with 
the large gravitational radius compared to the curvature scale in 
the bulk can be considered as a section of a black string~\cite{TKS}. 
Hence, it is important
to clarify how the gravitational waves are generated in the perturbed 
black string system and how the effects of the Gregory-Laflamme instability
 come into the observed signal of gravitational waves. 
For these purposes,  we need a basic formalism for black string perturbations
 in RS1 model.

In general relativity,  the perturbation around Kerr black 
hole is elegantly treated in the Newman-Penrose formalism~\cite{NP}.
Indeed, Teukolsky derived a separable master equation for the gravitational
waves in the Kerr black hole background~\cite{Teukolsky}. 
One purpose of this paper is to extend the Teukolsky
formalism to the braneworld context and derive the effective Teukolsky
equation~\cite{Kanno:2003au}. 

When the horizon size of the black string gets close to the Compton
 wavelength of the lowest Kaluza-Klein mode, the Gregory-Laflamme instability
 commences.  The other purpose of this paper is to explore this phenomena 
 from the brane point of view.
  
The organization of this paper is as follows.
In sec.II, we describe the black string in two-brane system and 
 highlight the role of Weyl tensor $E_{\mu\nu}$ in deriving
the effective Teukolsky equation. 
In sec.III, a master equation and junction conditions
 for $  E_{\mu\nu}$ are obtained.
  In sec.IV, we solve the master equation at low energy 
  using the gradient expansion 
  method and give the explicit solution for $ E_{\mu\nu}$. 
  Then, the effective Teukolsky  equation is obtained.
  In sec.V, we analyze the effect of the Gregory-Laflamme instability 
  on the braneworld. 
  The final section is devoted to the conclusion.
 
%===============================================================%
%************************ SECTION II ***************************%
%===============================================================%
\section{ Black String in RS1 and Teukolsky Equation}

Let us recall some basics of the black hole perturbations.
Based on the Newmann-Penrose (NP) null-tetrad formalism, in which the
tetrad components of the curvature tensor are the fundamental variables,
a master equation for the curvature perturbation was developed by Teukolsky
for a Kerr black hole with source. The master equation is called the Teukolsky
equation, and it is a wave equation for a null-tetrad component of the 
Weyl tensor $\Psi_0=-C_{pqrs}\ell^pm^q\ell^rm^s$ or $\Psi_4=-C_{pqrs}n^p\bar{m}^qn^r\bar{m}^s$, where $C_{pqrs}$ is the Weyl tensor and $\ell, n, m, \bar{m}$ 
are null basis in the NP formalism. All information about the 
gravitational radiation flux at infinity and at the event horizon can be 
extracted from $\Psi_0$ and $\Psi_4$. 
The Teukolsky equation is constructed by combining the Bianchi identity
with the Einstein equations. The Riemann tensor in the Bianchi identity
is written in terms of the Weyl tensor and the Ricci tensor. The Ricci
tensor is replaced by the matter fields using the Einstein equations.
In this way, the Bianchi identity becomes no longer identity and one
can get a master equation in which the curvature tensor is the fundamental 
variable~\cite{Chandra}.

It is of interest to extend the Teukolsky equation to the  braneworld. 
In the RS1 model, the two  3-branes are embedded in AdS$_5$ with the curvature
radius $\ell$ and the brane tensions given by
$\sigma_\oplus=6/(\kappa^2\ell)$ and $\sigma_\ominus=-6/(\kappa^2\ell)$. 
Our system is described by the action 
%%%%%%%%%%%%%%%
\begin{eqnarray}
S&=&\frac{1}{2 \kappa^2}\int d^5 x 
\sqrt{-\overset{(5)}{g}}\left({\cal R}
+\frac{12}{\ell^2}\right)
	-\sum_{i=\oplus,\ominus}\sigma_i 
	\int d^4 x \sqrt{-g^{i\mathrm{\hbox{-}brane}}} 
	+\sum_{i=\oplus,\ominus} 
	\int d^4 x \sqrt{-g^{i\mathrm{\hbox{-}brane}}}
	\,{\cal L}_{\rm matter}^i \ ,\label{5D:action}
\end{eqnarray}
%%%%%%%%%%%%%%%%
where $\overset{(5)}{g}_{\mu\nu}$, ${\cal R}$, 
$g^{i\mathrm{\hbox{-}brane}}_{\mu\nu}$, and $\kappa^2$ 
are the 5-dimensional metric, the 5-dimensional scalar curvature, 
the induced metric on the $i$-brane, 
and the 5-dimensional gravitational constant, respectively. 
It is easy to show the metric
\begin{eqnarray}
   ds^2 = dy^2 + e^{-2y/\ell} g_{\mu\nu} (x) dx^\mu dx^\nu
\end{eqnarray}
is the exact vacuum solution of the two-brane system
provided that the $g_{\mu\nu}$ satisfies the Ricci flat
 condition
\begin{eqnarray}
    R_{\mu\nu} (g) =0 \ .
\end{eqnarray}
Thus, a black hole on the brane is expected to be a section of black 
string.  
In fact,  Kerr metric leads to the exact solution for this model:   
\begin{eqnarray}
ds^2&=&dy^2+e^{-2\frac{y}{\ell}}\left[
	(1-2Mr/\Sigma)dt^2+(4Mar\sin^2\theta/\Sigma)dtd\varphi
	-(\Sigma/\triangle)dr^2-\Sigma d\theta^2
	\right.\nonumber\\
&&\qquad\qquad\qquad\left.
	-\sin^2\theta (r^2+a^2+2Ma^2r\sin^2\theta/\Sigma)d\varphi^2)
	\right]  \ ,
	\label{Kerr BS}
\end{eqnarray}
where $y$ is the coordinate of extra-dimension, $M$ is the mass of the black 
hole, $aM$ its angular momentum, $\Sigma=r^2+a^2\cos^2\theta$ and $\triangle=
r^2-2Mr+a^2$. When $a=0$, the metric reduces to the Schwarzschild metric on 
the brane.

Since the Bianchi identity is
independent of dimensions, what we need is the projected Einstein equations
on the brane derived by Shiromizu, Maeda and Sasaki~\cite{ShiMaSa}. 
The first order perturbation of the projected Einstein equation is 
\begin{eqnarray}
G_{\mu\nu}=8\pi GT_{\mu\nu}- \delta E_{\mu\nu} \ ,
\label{SMS}
\end{eqnarray}
where $8\pi G =\kappa^2 /\ell$.
If we replace the Ricci curvature in the Bianchi identity to
the matter fields and $E_{\mu\nu}$ using Eq.~(\ref{SMS}), then the projected 
Teukolsky equation on the brane is written in the following form,
\begin{eqnarray}
&&[(\triangle+3\gamma-\gamma^{\ast}+4\mu+\mu^{\ast})
	(D+4\epsilon-\rho)
	-(\delta^{\ast}-\tau^{\ast}+\beta^{\ast}+3\alpha+4\pi)
	(\delta-\tau+4\beta)-3 \Psi_2]\delta \Psi_4 \nonumber\\
&&\quad	={1\over 2} (\triangle+3\gamma-\gamma^{\ast}+4\mu+\mu^{\ast})
	[(\delta^{\ast}-2\tau^{\ast}+2\alpha)
	(8\pi G T_{nm^{\ast}}- \delta E_{nm^{\ast}}) \nonumber \\
&&\hspace{6cm}	
	-(\triangle+2\gamma-2\gamma^{\ast}+\mu^{\ast})
	(8\pi G T_{m^{\ast}m^{\ast}}
	- \delta E_{m^{\ast}m^{\ast}})]\nonumber\\&&\qquad 
	+{1\over 2}(\delta^{\ast}-\tau^{\ast}+\beta^{\ast}+3\alpha+4\pi)
	[(\triangle+2\gamma+2\mu^{\ast})(8\pi G T_{nm^{\ast}}
	- \delta E_{nm^{\ast}}) \nonumber \\
&&\hspace{6cm}
	-(\delta^{\ast}-\tau^{\ast}+2\beta^{\ast}+2\alpha)
	(8\pi G T_{nn}- \delta E_{nn})] \ .
	\label{teukolsky}
\end{eqnarray}
Here our notation follows that of  \cite{Teukolsky}.
We see the effects of a fifth dimension, $\delta E_{\mu\nu}$, is described as
a source term in the projected Teukolsky equation. It should be stressed that
 the projected Teukolsky equation on the brane Eq.~(\ref{teukolsky})
is not a closed system yet. One must solve the gravitational field in the bulk
 to obtain $\delta E_{\mu\nu}$. 

%===============================================================%
%************************ SECTION II ***************************%
%===============================================================%

\section{Formalism}

In this section, we will derive the basic equations needed to analyze
the gravitational waves emitted from the black string.

 Firstly, we write down 5-dimensional Einstein equations and
 the junction conditions at the brane positions. 
 Let $n$ be a unit normal vector field to branes. 
Using the extrinsic curvature $K_{\mu\nu}=-(1/2)\mbox \pounds_n g_{\mu\nu}$, 
 5-dimensional Einstein equations in the bulk become
\begin{eqnarray}
&&-\frac{1}{2}\overset{(4)}{R}+\frac{1}{2}K^2
	-\frac{1}{2}K^{\alpha\beta}K_{\alpha\beta}
	=\frac{6}{\ell^2} \ ,
	\label{5D-Einstein:yy}\\
&&K_{\mu}{}^{\lambda}{}_{|\lambda}-K_{|\mu}=0 \ ,
	\label{5D-Einstein:ymu}\\
&&  \overset{(4)}{G}{}^{\mu}{}_{\nu} = -E^{\mu}{}_{\nu}
	+\frac{3}{\ell^2}\delta^{\mu}_{\nu}
	-K^{\mu\alpha}K_{\alpha\nu}+KK^{\mu}{}_{\nu}
	+\frac{1}{2}\delta^{\mu}_{\nu}
	\left(K^{\alpha\beta}K_{\alpha\beta}-K^2\right) \ ,
	\label{5D-Einstein:munu}
\end{eqnarray}
where ``the electric part" of the Weyl tensor
\begin{eqnarray}
E_{\mu\nu}
	=\mbox \pounds_nK_{\mu\nu}+K_{\mu}{}^{\alpha}K_{\alpha\nu}
	+\frac{1}{\ell^2}g_{\mu\nu}
	\label{weyl:electric}
\end{eqnarray}
is defined. Here, (4) represents the 4-dimensional quantity and
 $|$ is the covariant derivative with respect to the metric 
$g_{\mu\nu}(y,x^\mu)$.
 As the branes act as the singular sources, 
 we also have the junction conditions 
\begin{eqnarray}
&&\left[K^{\mu}{}_{\nu}-\delta^{\mu}_{\nu}K\right]\Big|_{\oplus}
	=\frac{\kappa^2}{2}\left(-\overset{\oplus}{\sigma}\delta^{\mu}_{\nu}
	+\overset{\oplus}{T}{}^{\mu}{}_{\nu}\right) \ ,
	\label{JC-k:p}\\
&&\left[K^{\mu}{}_{\nu}-\delta^{\mu}_{\nu}K\right]\Big|_{\ominus}
	=-\frac{\kappa^2}{2}\left(-\overset{\ominus}{\sigma}\delta^{\mu}_{\nu}
	+\overset{\ominus}{T}{}^{\mu}{}_{\nu}\right) \ .
	\label{JC-k:n}
\end{eqnarray}

 To obtain a master equation for $\delta E_{\mu\nu}$, 
 we start with the 5-dimensional Bianchi identities. 
 Using the Gauss equation and the Codacci equation, we obtain
\begin{eqnarray}
&&\hspace{-5mm}
\mbox \pounds_nB_{\mu\nu\lambda}+E_{\mu\nu|\lambda}-E_{\mu\lambda|\nu}
	+K^{\alpha}{}_{\mu}B_{\alpha\nu\lambda}
	+K^{\alpha}{}_{\lambda}B_{\nu\alpha\mu}
	-K^{\alpha}{}_{\nu}B_{\lambda\alpha\mu}
	=0 \ ,
	\label{bianchi1}\\
&&\hspace{-5mm}
B_{\mu[\nu\lambda|\rho]}
	+K^{\sigma}{}_{[\rho}\overset{(4)}{R}_{\nu\lambda]\sigma\mu}
	=0 \ ,
	\label{bianchi2}\\
&&\hspace{-5mm}
\mbox \pounds_n\overset{(4)}{R}_{\mu\nu\lambda\rho}
	+K^{\alpha}{}_{\mu}\overset{(4)}{R}_{\alpha\nu\lambda\rho}
	-K^{\alpha}{}_{\nu}\overset{(4)}{R}_{\alpha\mu\lambda\rho}
	+B_{\lambda\mu\nu|\rho}
	-B_{\rho\mu\nu|\lambda}
	=0 \ ,
	\label{bianchi3}\\
&&\hspace{-5mm}
\overset{(4)}{R}_{\mu\nu[\lambda\rho|\sigma]}=0 \ ,
	\label{bianchi4}
\end{eqnarray}
where we have defined ``the magnetic part" of the Weyl tensor
\begin{eqnarray}
B_{\mu\nu\lambda}= K_{\mu\lambda|\nu}-K_{\mu\nu|\lambda} \ .
	\label{weyl:magnetic}
\end{eqnarray}
After combining (\ref{5D-Einstein:yy}) with (\ref{5D-Einstein:munu}) and
putting the result into Eq.~(\ref{bianchi3}), we have
\begin{eqnarray}
\mbox \pounds_nE_{\mu\nu}&=&B_{(\mu\nu)\rho}{}^{|\rho}
	-\Sigma^{\alpha\beta}\overset{(4)}{C}_{\mu\alpha\nu\beta}
        +{1\over 2} g_{\mu\nu} E_{\alpha \beta} \Sigma^{\alpha \beta}
	-2 ( \Sigma^{\alpha}{}_{\mu}E_{\alpha\nu}
	+\Sigma^{\alpha}{}_{\nu}E_{\alpha\mu})\nonumber\\
&&	+{1\over 2} K E_{\mu\nu}
	+{1\over 2} g_{\mu\nu} \Sigma_\alpha{}^\beta
	\Sigma_\beta{}^\gamma \Sigma_\gamma{}^{\alpha} 
	- 2 \Sigma_{\mu}{}^{\alpha}\Sigma_{\alpha}{}^{\beta}\Sigma_{\beta\nu}
	+{7\over 6} \Sigma_{\mu\nu} \Sigma^\alpha{}_\beta 
	\Sigma^\beta{}_\alpha  \ ,
	\label{eq:weyl}
\end{eqnarray}
where we decomposed the extrinsic curvature into the traceless part
and the trace part
\begin{eqnarray}
K_{\mu\nu}=\Sigma_{\mu\nu}
	+\frac{1}{4}g_{\mu\nu}K \ .
\end{eqnarray}

Now we consider the perturbation of these equations. 
We generalize the background metric Eq.~(\ref{Kerr BS}) to
a Ricci flat string without source 
($T_{\mu\nu}=0$) whose metric is written as
\begin{eqnarray}
ds^2=dy^2+e^{-2\frac{y}{\ell}}g_{\mu\nu}(x^\mu)dx^\mu dx^\nu
\label{metric:BS} \ ,
\end{eqnarray}
where $g_{\mu\nu}(x^\mu)$ is supposed to be the Ricci flat metric.
 In the case of Kerr metric, the metric (\ref{metric:BS}) 
 represents the rotating
 black string~\cite{Modgil}. 
 The variables $K_{\mu\nu}$, $E_{\mu\nu}$ and $B_{\mu\nu}$ for this
background are given by
\begin{eqnarray}
K_{\mu\nu}=\frac{1}{\ell}g_{\mu\nu} 
	\ , \qquad
	E_{\mu\nu}=B_{\mu\nu\lambda}=0 \ .
\end{eqnarray}
Linearizing  Eq. (\ref{eq:weyl}) around this background, we obtain
\begin{eqnarray}
\delta E_{\mu\nu,y} = \delta B_{(\mu\nu)\rho}^{|\rho}
	-\overset{(4)}{C}_{\mu\alpha\nu\beta} \delta \Sigma^{\alpha\beta} 
	+{2\over \ell } \delta E_{\mu\nu}
  	\label{ptb:weyl1}\ .
\end{eqnarray}
Similarly,  Eq.~(\ref{bianchi1}) reduces to
\begin{eqnarray}
\delta B_{(\mu\nu)\alpha}{}^{|\alpha}=
       -\delta E_{\mu\nu|\lambda} + \delta E_{\mu\lambda|\nu} \ . 
	\label{eq1}
\end{eqnarray}
 Using the following relation
\begin{eqnarray}
  \left( \delta B_{(\mu\nu)\lambda,y} \right)^{|\lambda}
  = \left( \delta B_{(\mu\nu)\lambda}{}^{|\lambda} \right)_{,y}
  -{2\over \ell} \delta B_{(\mu\nu)\lambda}{}^{|\lambda} \ ,
\end{eqnarray}
we can eliminate $B_{\mu\nu\lambda}$ from Eqs.~(\ref{ptb:weyl1})
 and (\ref{eq1}).
Thus, the master equation  for $\delta E_{\mu\nu}$  is found as
\begin{eqnarray}
\left(\partial^2_y-\frac{4}{\ell}\partial_{y}+\frac{4}{\ell^2}\right)
	\delta E_{\mu\nu}&=& 
	-e^{2\frac{y}{\ell}}{\hat{\cal L}_{\mu\nu}{}^{\alpha\beta}}
	\delta E_{\alpha\beta} \nonumber \\
     &\equiv& - e^{2\frac{y}{\ell}}\hat{\cal L}\delta E_{\mu\nu} \ ,
	\label{eom:weyl}
\end{eqnarray}
where ${\hat{\cal L}_{\mu\nu}{}^{\alpha\beta}}
	= \Box \delta_\mu^\alpha \delta_\nu^\beta 
	+2 R_\mu{}^\alpha{}_\nu{}^\beta $
stands for the Lichnerowicz operator. 
Here, the covariant derivative and the Riemann tensor are constructed 
 from $g_{\mu\nu} (x)$. 

In order to deduce the junction conditions for $\delta E_{\mu\nu}$
 from Eqs.~(\ref{JC-k:p}) and (\ref{JC-k:n}), 
 we use Eq.~(\ref{ptb:weyl1}) and
\begin{eqnarray}
\delta B_{\mu\nu\rho}=\delta K_{\mu\rho|\nu}
	-\delta K_{\mu\nu|\rho} \ .
\end{eqnarray}
As a result, the junction conditions on each branes become
\begin{eqnarray}
&& e^{2\frac{y}{\ell}}\left[e^{-2\frac{y}{\ell}}\delta E_{\mu\nu}\right]_{,y}
\bigg{|}_{y=0} =-\frac{\kappa^2}{6}\overset{\oplus}{T}_{|\mu\nu}
	-\frac{\kappa^2}{2}{\hat{\cal L}_{\mu\nu}{}^{\alpha\beta}}
	\left( \overset{\oplus}{T}_{\alpha\beta}
	-\frac{1}{3}g_{\alpha\beta}\overset{\oplus}{T}\right) \ ,
	\label{JC:p-1}\ \\
&&e^{2\frac{y}{\ell}}\left[e^{-2\frac{y}{\ell}}\delta E_{\mu\nu}\right]_{,y}
\bigg{|}_{y=d}=\frac{\kappa^2}{6}\overset{\ominus}{T}_{|\mu\nu}
	+\frac{\kappa^2}{2}{\hat{\cal L}_{\mu\nu}{}^{\alpha\beta}}
	\left( \overset{\ominus}{T}_{\alpha\beta}
	-\frac{1}{3}g_{\alpha\beta}\overset{\ominus}{T}\right) \ .
	\label{JC:n-1}
\end{eqnarray}
Let $\overset{\oplus}{\phi}(x)$ and $\overset{\ominus}{\phi}(x)$
be the scalar fields on each branes which satisfy
\begin{eqnarray}
\Box \overset{\oplus}{\phi}&=&{\kappa^2\over 6} 
	\overset{\oplus}{T} \ ,
	\label{radion1} \\
\Box \overset{\ominus}{\phi}&=&{\kappa^2\over 6} 
	\overset{\ominus}{T} \ ,
	\label{radion2}
\end{eqnarray}
respectively \cite{Gen}. In the Ricci flat space-time, the identity
\begin{eqnarray}
\left(\Box\phi\right)_{|\mu\nu}
	={\hat{\cal L}_{\mu\nu}{}^{\alpha\beta}}
	\left(\phi_{|\alpha\beta}\right)
	\label{relation}
\end{eqnarray}
holds. Thus, Eqs.(\ref{JC:p-1}) and (\ref{JC:n-1}) can be rewritten as
\begin{eqnarray}
&&\hspace{-5mm}
e^{2\frac{y}{\ell}}
	\left[
	e^{-2\frac{y}{\ell}}\delta E_{\mu\nu}
	\right]_{,y}
	\bigg{|}_{y=0}
	=- \hat{\cal L} \left(\overset{\oplus}{\phi}_{|\mu\nu}
	- g_{\mu\nu} \Box \overset{\oplus}{\phi}\right)
	-\frac{\kappa^2}{2} \hat{\cal L}
	 \overset{\oplus}{T}_{\mu\nu}
	~\equiv \hat{\cal L} \overset{\oplus}{S}_{\mu\nu} \ ,
	\label{JC:p}\ \\
&&\hspace{-5mm}
e^{2\frac{y}{\ell}}
	\left[
	e^{-2\frac{y}{\ell}}\delta E_{\mu\nu}
	\right]_{,y}
	\bigg{|}_{y=d}
	=\hat{\cal L}\left(\overset{\ominus}{\phi}_{|\alpha\beta}
	- g_{\mu\nu} \Box \overset{\ominus}{\phi} \right)
	+\frac{\kappa^2}{2} \hat{\cal L}
	 \overset{\ominus}{T}_{\mu\nu}
	~\equiv - \hat{\cal L} \overset{\ominus}{S}_{\mu\nu} \ .
	\label{JC:n}
\end{eqnarray}
The scalar fields $\overset{\oplus}{\phi}$ and $\overset{\ominus}{\phi}$
 can be interpreted as the brane fluctuation modes.
 
%===============================================================%
%************************* SECTION IV **************************%
%===============================================================%
\section{Effective Teukolsky Equation At Low Energy}

\subsection{Gradient Expansion Method}

 It is known that the Gregory-Laflamme instability occurs if 
 the curvature length scale of the black hole $L$ is less than the Compton
 wavelength of Kaluza-Klein (KK) modes $\sim \ell \exp(d/\ell) $~\cite{GL}.  
 As we are interested in the stable rotating black string, 
\begin{equation}
\epsilon~=~ \left({\ell \over L}\right)^2 \ll 1 
\end{equation}
 is assumed. This means that the curvature on the brane can be neglected 
compared with the derivative with respect to $y$. 
 Our iteration scheme consists in writing the Weyl tensor $E_{\mu\nu}$
 in  the order of $\epsilon$~\cite{Kanno}. 
Hence, we will seek the Weyl tensor as a perturbative series
\begin{eqnarray}
&&\hspace{-3mm}
\delta E_{\mu\nu}(y,x^\mu )
	=\delta\overset{(1)}{E}_{\mu\nu} (y,x^\mu)
	+\delta\overset{(2)}{E}_{\mu\nu}(y, x^\mu )
	+\delta\overset{(3)}{E}_{\mu\nu}(y, x^\mu ) + \cdots  \ .
\end{eqnarray}
%

%===============================================================%
%************************ SUB SECTION **************************%
%===============================================================%
\subsubsection{First order}
At first order, we can neglect the Lichnerowicz operator term. 
Then Eq.~(\ref{eom:weyl}) become
\begin{eqnarray}
\left(\partial^2_y-\frac{4}{\ell}\partial_{y}+\frac{4}{\ell^2}\right)
	\delta\overset{(1)}{E}{}_{\mu\nu}=0 \ ,
	\label{eom:weyl1}
\end{eqnarray}
where the superscript (1) represents the order of the derivative expansion.
This can be readily integrated into
\begin{eqnarray}
\delta\overset{(1)}{E}_{\mu\nu}=e^{2\frac{y}{\ell}}\left(
	\overset{(1)}{C}_{\mu\nu}\frac{y}{\ell}
	+\overset{(1)}{\chi}_{\mu\nu}\right) \ ,
	\label{1:weyl1}
\end{eqnarray}
where $\overset{(1)}{C}_{\mu\nu}$ and $\overset{(1)}{\chi}_{\mu\nu}$
are the constants of integration which depend only on $x^\mu$ and satisfy 
the transverse 
$\overset{(1)}{C}{}^{\mu}{}_{\nu|\mu}
=\overset{(1)}{\chi}{}^{\mu}{}_{\nu|\mu}=0$ and traceless 
$\overset{(1)}{C}{}^{\mu}{}_{\mu}
=\overset{(1)}{\chi}{}^{\mu}{}_{\mu}=0$ constraints. 
The junction conditions on each branes at this order are 
\begin{eqnarray}
e^{2\frac{y}{\ell}}\left[e^{-2\frac{y}{\ell}} 
\delta\overset{(1)}{E}{}_{\mu\nu}\right]_{,y}
\Bigg{|}_{y=0,d} = 0 \ .
\label{JC1:pn} 
\end{eqnarray}
Imposing this junction condition 
Eq.~(\ref{JC1:pn}) on the solution (\ref{1:weyl1}), we see 
$\overset{(1)}{C}_{\mu\nu}=0$. 
 Thus, we get the first order Weyl tensor 
\begin{eqnarray}
\delta\overset{(1)}{E}_{\mu\nu}=e^{2\frac{y}{\ell}}
	\overset{(1)}{\chi}{}_{\mu\nu}(x)
	\label{1:weyl2}\ ,
\end{eqnarray}
where $\overset{(1)}{\chi}{}_{\mu\nu}(x)$ is arbitrary at this order.
 This should be determined from the next order analysis. 

%===============================================================%
%************************ SUB SECTION **************************%
%===============================================================%
\subsubsection{Second order}

The next order solutions are obtained by taking into account the
terms neglected at first order. At second order, Eq.~(\ref{eom:weyl}) becomes
\begin{eqnarray}
\left(\partial^2_y-\frac{4}{\ell}\partial_{y}+\frac{4}{\ell^2}\right)
	\delta \overset{(2)}{E}{}_{\mu\nu}=-e^{2\frac{y}{\ell}}
	\hat{\cal L} \delta \overset{(1)}{E}{}_{\mu\nu} \ .
	\label{eom:weyl2}
\end{eqnarray}
Substituting the first order $\overset{(1)}{E}{}_{\alpha\beta}$ into the right
hand side of Eq.~(\ref{eom:weyl2}), we obtain
\begin{eqnarray}
 \delta \overset{(2)}{E}{}_{\mu\nu}=
	e^{2\frac{y}{\ell}} \left(
	\overset{(2)}{C}_{\mu\nu}~\frac{y}{\ell} 
	+\overset{(2)}{\chi}_{\mu\nu}\right)
	-{\ell^2 \over 4}e^{4\frac{y}{\ell}}
  	\hat{\cal L} \overset{(1)}{\chi}_{\mu\nu} \ , 
  	\label{2:weyl2}
\end{eqnarray}
where $\overset{(2)}{C}_{\mu\nu}$ and $\overset{(2)}{\chi}_{\mu\nu}$
are again the  constants of integration at this order and 
satisfy the transverse and traceless constraint 
($\overset{(2)}{\chi}{}^{\mu}{}_{\mu}
=\overset{(2)}{\chi}{}^{\mu}{}_{\nu|\mu}=0$, etc.). The junction conditions 
at this order give
\begin{eqnarray}	
&&\left[e^{-2\frac{y}{\ell}}
	\delta \overset{(2)}{E}{}_{\mu\nu}\right]_{,y}
	\Bigg{|}_{y=0}= \hat{\cal L} \overset{\oplus}{S}_{\mu\nu}{}^{(1)} \ ,
	\label{JC2:p}
	\ \\
&&\left[e^{-2\frac{y}{\ell}}
	\delta \overset{(2)}{E}{}_{\mu\nu}\right]_{,y}
	\Bigg{|}_{y=d}=- \Omega^2 \hat{\cal L} 
	\overset{\ominus}{S}_{\mu\nu}{}^{(1)} \ .
	\label{JC2:n}
\end{eqnarray}
Here $\Omega^2=\exp[-2\frac{d}{\ell}]$ is a conformal factor that relates 
the metric on the $\oplus$-brane to that on the $\ominus$-brane~\cite{Kanno}. 
Substituting Eq.~(\ref{2:weyl2}) into the above junction conditions, we get 
\begin{eqnarray}	
&&\frac{1}{\ell}\overset{(2)}{C}_{\mu\nu}
	-\frac{\ell}{2} \hat{\cal L}
  	\overset{(1)}{\chi}_{\mu\nu}
	=\hat{\cal L} \overset{\oplus}{S}_{\mu\nu}{}^{(1)} \ ,
	\label{JC2:p}
	\ \\
	&&
\frac{1}{\ell}\overset{(2)}{C}_{\mu\nu}
	-\frac{\ell}{2}~\frac{1}{\Omega^2}
	\hat{\cal L} \overset{(1)}{\chi}_{\mu\nu}
  	=- \Omega^2 \hat{\cal L} \overset{\ominus}{S}_{\mu\nu}{}^{(1)} \ .
	\label{JC2:n}
\end{eqnarray}
Eliminating $\overset{(1)}{\chi}_{\alpha\beta}$ from these equations,
we obtain one of the constants of integration
\begin{eqnarray}
\overset{(2)}{C}_{\mu\nu}
	&=&-{\ell\over 1-\Omega^2}
	\left( \hat{\cal L} \overset{\oplus}{S}_{\mu\nu}^{(1)}
    + \Omega^4 \hat{\cal L} \overset{\ominus}{S}_{\mu\nu}^{(1)} \right) \ .
\end{eqnarray}
Similarly, eliminating $\overset{(2)}{C}_{\mu\nu}$ from Eqs.(\ref{JC2:p}) and 
(\ref{JC2:n}),  we obtain the equation
\begin{eqnarray}
\hat{\cal L}\overset{(1)}{\chi}_{\mu\nu}
	&=&{2 \over \ell}{\Omega^2 \over 1-\Omega^2}
	\left(  \hat{\cal L} \overset{\oplus}{S}_{\mu\nu}^{(1)}
	+\Omega^2 \hat{\cal L} \overset{\ominus}{S}_{\mu\nu}^{(1)} \right) \ ,
\end{eqnarray}
which is easily integrated as
\begin{eqnarray}
\overset{(1)}{\chi}_{\mu\nu}
	={2 \over \ell}{\Omega^2 \over 1-\Omega^2}
          \left(  \overset{\oplus}{S}_{\mu\nu}^{(1)}
	     + \Omega^2  \overset{\ominus}{S}_{\mu\nu}^{(1)} \right) \ .
\end{eqnarray}
 Comparing Eq.(46) with the analysis for the perturbations around the 
 flat two-brane background, we see the above result corresponds to
  the zero mode contribution~\cite{Kanno}. Hence, the KK corrections
  come from the second order corrections which are not yet determined.

%===============================================================%
%************************ SUB SECTION **************************%
%===============================================================%
\subsubsection{Third order}

 In order to obtain KK corrections, we need to fix 
 $\overset{(2)}{\chi}_{\mu\nu}$. For that purpose, we must proceed to third
 order analysis.  At third order, we have the solution 
\begin{eqnarray}
\delta \overset{(3)}{E}{}_{\mu\nu}
	&=&
	e^{2\frac{y}{\ell}}
	\left(
	\overset{(3)}C_{\mu\nu}~{y\over \ell}+\overset{(3)}\chi_{\mu\nu}
	\right) 
	-\frac{\ell^2}{4}e^{4\frac{y}{\ell}}
         \hat{\cal L} \overset{(2)}{\chi}_{\mu\nu}
	-\frac{\ell}{4}(y-\ell) e^{4\frac{y}{\ell}}
        \hat{\cal L} \overset{(2)}{C}_{\mu\nu} 
	+  \frac{\ell^4}{64}e^{6 \frac{y}{\ell}}
	\hat{\cal L}^2 \overset{(1)}{\chi}_{\mu\nu}      \ , 
\end{eqnarray}
where $\overset{(3)}C_{\mu\nu}$ and $\overset{(3)}\chi_{\mu\nu}$ are the
 constants of integration at this order. 
 Junction conditions yield
\begin{eqnarray}
&&    \frac{\ell^3}{16} \hat{\cal L}^2
	\overset{(1)}{\chi}_{\mu\nu} 
	+\frac{\ell}{4} \hat{\cal L} \overset{(2)}{C}_{\mu\nu} 
      -\frac{\ell}{2} \hat{\cal L}
	\overset{(2)}{\chi}_{\mu\nu}
       +{1\over \ell} \overset{(3)}{C}_{\mu\nu}
	= \hat{\cal L} \overset{\oplus}{S}_{\mu\nu} \ , \qquad \\
&& -\frac{1}{2\Omega^2}(d-{\ell \over 2})
          \hat{\cal L} \overset{(2)}{C}_{\mu\nu} 
	+  \frac{\ell^3}{16\Omega^4}  \hat{\cal L}^2
	\overset{(1)}{\chi}_{\mu\nu} 
	-\frac{\ell}{2 \Omega^2} \hat{\cal L}
	\overset{(2)}{\chi}_{\mu\nu} 
	+ {1\over \ell} \overset{(3)}C_{\mu\nu} 
	= -\Omega^2 \hat{\cal L} \overset{\ominus}{S}_{\mu\nu} \ .
\end{eqnarray}
 We get $\overset{(3)}{C}_{\mu\nu}$ 
 from above equations as
\begin{eqnarray}
\overset{(3)}{C}_{\mu\nu}
	&=& -  {\ell d\over 2} {1\over 1- \Omega^2 } 
	\hat{\cal L} \overset{(2)}{C}_{\mu\nu} 
	+ {\ell^4 \over 16}  {1\over \Omega^2} 
	\hat{\cal L}^2 \overset{(1)}{\chi}_{\mu\nu}
	+ {\ell \over 1-\Omega^2} 
	\left[ \hat{\cal L} \overset{\oplus}{S}{}^{(2)}_{\mu\nu} 
	+ \Omega^4 \hat{\cal L} \overset{\ominus}{S}
	{}^{(2)}_{\mu\nu} \right] 
\end{eqnarray}
and
\begin{eqnarray}
 \hat{\cal L}\overset{(2)}{\chi}_{\mu\nu}
	&=&
	\left[ {1\over 2} + {d\over \ell} {1\over \Omega^2 -1} \right]
	\hat{\cal L} \overset{(2)}{C}_{\mu\nu}
	+ {\ell^2 \over 8} \left( 1+ {1\over \Omega^2} \right)
	\hat{\cal L}^2 \overset{(1)}{\chi}_{\mu\nu}
	+{2\over \ell} {\Omega^2 \over 1-\Omega^2} 
	\left[ \hat{\cal L} \overset{\oplus}{S}{}^{(2)}_{\mu\nu} 
	+ \Omega^2 \hat{\cal L} \overset{\ominus}{S}
	{}^{(2)}_{\mu\nu} \right] \ .
\end{eqnarray}
It is easy to obtain 
\begin{eqnarray}
\overset{(2)}{\chi}_{\mu\nu}
	&=&\left[ {1\over 2} + {d\over \ell} {1\over \Omega^2 -1} \right]
	 \overset{(2)}{C}_{\mu\nu}
	 + {\ell^2 \over 8} \left( 1+ {1\over \Omega^2} \right)
	\hat{\cal L} \overset{(1)}{\chi}_{\mu\nu}
	+{2\over \ell} {\Omega^2 \over 1-\Omega^2} 
	\left[  \overset{\oplus}{S}{}^{(2)}_{\mu\nu} 
	+ \Omega^2  \overset{\ominus}{S}
	{}^{(2)}_{\mu\nu} \right] \ .
\end{eqnarray}
Thus, we have obtained KK corrections.
 In principle, we can continue this perturbative calculations
 to any order.

\subsection{Effective Teukolsky Equation}

Schematically, Teukolsky equation (9) takes the following form
\begin{eqnarray}
    \hat{P} \delta \Psi_4 
    = \hat{Q} \left( 8\pi G T_{nm^{\ast}} - \delta E_{nm^{\ast}} \right) 
    + \cdots \ ,
\end{eqnarray}
where $\hat{P}$ and $\hat{Q}$ are the operators  in Eq.(9). 
 What we needed is $\delta E_{\mu\nu}$ in the above equation. 
Now, we can write down $\delta E_{\mu\nu}$ on the brane up to the second order
 as
\begin{eqnarray}
\delta E_{\mu\nu}\Big|_{y=0} &=& {2\over \ell} {\Omega^2 \over 1-\Omega^2} 
	\left[  \overset{\oplus}{S}{}_{\mu\nu} 
	+ \Omega^2  \overset{\ominus}{S}
	{}_{\mu\nu} \right]
	+ \left[ {1\over 2} + {d\over \ell} {1\over \Omega^2 -1} \right]
	{\ell \over 1-\Omega^2} 
	\left[ \hat{\cal L} \overset{\oplus}{S}{}_{\mu\nu} 
	+ \Omega^4 \hat{\cal L} \overset{\ominus}{S}
	{}_{\mu\nu} \right] \nonumber \\
&&	+  {\ell \over 4} 
	\left[ \hat{\cal L} \overset{\oplus}{S}{}_{\mu\nu} 
	+ \Omega^2 \hat{\cal L} \overset{\ominus}{S}
	{}_{\mu\nu} \right] 
	\label{extrasource}\ ,
\end{eqnarray}
where
\begin{eqnarray}
\overset{\oplus}{S}_{\mu\nu}
    &=&   - \overset{\oplus}{\phi}_{|\mu\nu}
         + g_{\mu\nu} \Box \overset{\oplus}{\phi}
	-\frac{\kappa^2}{2} \overset{\oplus}{T}_{\mu\nu}
	\ , 
                     \\
\overset{\ominus}{S}_{\mu\nu}
    &=& - \overset{\ominus}{\phi}_{|\mu\nu}
          + g_{\mu\nu} \Box \overset{\ominus}{\phi}
	-\frac{\kappa^2}{2} \overset{\ominus}{T}_{\mu\nu}
	\ . 
\end{eqnarray}
Substituting this $\delta E_{\mu\nu}$ into (\ref{teukolsky}), 
we get the effective
 Teukolsky equation on the brane. 
  From Eq.~(\ref{extrasource}), we see KK corrections give extra sources
 to Teukolsky equation. To obtain quantitative results, we must resort
  to numerical calculations. It should be stressed that the effective
 Teukolsky equation is separable like as the conventional Teukolsky equation.
 Therefore, it is suitable for numerical treatment. 

%===============================================================%
%************************* SECTION V ***************************%
%===============================================================%
\section{Gregory-Laflamme Instability:View from the Brane}

 It is  possible to apply our master equation (\ref{eom:weyl}) to
 the transition phenomena from black string to black hole.
 Interestingly,  our master equation coincides with the 
 5-dimensional Einstein equations in the harmonic gauge. Hence,
 it must exhibit the Gregory-Laflamme instability.  It would be interesting 
 to see  this instability from the  brane point of view. 

The Gregory-Laflamme instability takes place in the regime
$\epsilon\approx 1$ where the low energy gradient expansion method
cannot be applied. Hence, we will examine qualitative features of 
this phenomena through a formal Green function representation of the solution.
To solve  equation for $\delta E_{\mu\nu}$, we introduce the Green function
\begin{eqnarray}
&&\left[\left(\partial^2_y-\frac{4}{\ell}\partial_{y}+\frac{4}{\ell^2}\right)
	\delta^{\alpha}_{\mu}\delta^{\beta}_{\nu}
	+e^{2\frac{y}{\ell}}{\hat{\cal L}_{\mu\nu}{}^{\alpha\beta}}\right]
	G_{\alpha\beta}{}^{\lambda\rho}(x,y;x',y') \nonumber\\
&&\hspace{4cm}
	=-\frac{e^{4\frac{y}{\ell}}}{\sqrt{-g}}\delta^4(x-x')\delta(y-y')
	\left(\delta^{(\alpha}_{\mu}\delta^{\rho)}_{\nu}
	-\frac{1}{4}g_{\mu\nu}g^{\lambda\rho}\right)	
\end{eqnarray}
with the boundary conditions
\begin{eqnarray}
\partial_y\left[e^{-2\frac{y}{\ell}}G_{\mu\nu}{}^{\alpha\beta}\right]
	\bigg{|}_{y=0,d}=0 \ .
\end{eqnarray}
Then, the formal solution is given by
\begin{eqnarray}
&& \delta E_{\mu\nu}(x,y)
	=\int d^4x'\sqrt{-g} 
	\left[e^{-2\frac{y}{\ell}}G_{\mu\nu}{}^{\lambda\rho}
	\partial_y\left(e^{-2\frac{y}{\ell}} \delta E_{\lambda\rho}\right)
	\right]\bigg{|}_{y=0}^{y=d} \ .    \ \ \ \ \ \ 
\end{eqnarray}
Using junction conditions (\ref{JC:p}) and (\ref{JC:n}),
we obtain
\begin{eqnarray}
&& \delta E_{\mu\nu}(x',y')
	=\int d^4x\sqrt{-g} 
	\left[
	G_{\mu\nu}{}^{\lambda\rho}(x,d;x',y')\Omega^4 
	\hat{\cal L} \overset{\ominus}{S}_{\lambda \rho}
	+G_{\mu\nu}{}^{\lambda\rho}(x,0;x',y') 
	\hat{\cal L} \overset{\oplus}{S}_{\lambda\rho}
	\right]  \ ,
	\label{green1}
\end{eqnarray}
which can be rewritten as
\begin{eqnarray}
&&\hspace{-5mm}
	\delta E_{\mu\nu}(x',y')
	=\int d^4x\sqrt{-g} 
	\left[
	\hat{\cal L}_{\lambda\rho}{}^{\alpha\beta}G_{\mu\nu}{}^{\lambda\rho}
	(x,d;x',y')\Omega^4 
	\overset{\ominus}{S}_{\lambda \rho}
	+\hat{\cal L}_{\lambda\rho}{}^{\alpha\beta}G_{\mu\nu}{}^{\lambda\rho}
	(x,0;x',y') 
	\overset{\oplus}{S}_{\lambda\rho}
	\right]  \ .
	\label{green2}
\end{eqnarray}
where we dropped the surface term. If we use the 4-dimensional Green function  
\begin{eqnarray}
&&  \left[\hat{\cal L}_{\mu\nu}{}^{\alpha\beta}
	-\lambda^2_n\delta^{\alpha}_{\mu}\delta^{\beta}_{\nu}
	\right] 
	\overset{(4)}{G}(x,x';\lambda^2_n)_{\alpha\beta}{}^{\lambda\rho}
	= - {\delta (x-x') \over \sqrt{-g}} 
  \delta_\mu^\lambda \delta_\nu^\rho 
  \label{4d:green}\ ,
\end{eqnarray}
 the Green function can be expressed  as
\begin{eqnarray}
G_{\mu\nu}{}^{\alpha\beta}(x,y;x',y')=
	\sum\limits_n
	\varphi_n(y)\varphi_n(y')
 	\overset{(4)}{G}(x,x';\lambda^2_n)_{\mu\nu}{}^{\alpha\beta} \ .\ \ \ \
 	\label{5:4} 
\end{eqnarray}
Here, a mode function satisfies
\begin{eqnarray}
e^{-2\frac{y}{\ell}}\left(\partial_y-\frac{2}{\ell}\right)^2\varphi_n(y)
	=-\lambda_n^2\varphi_n(y) \ .
\end{eqnarray}
where $\lambda_n^2$ is an eigenvalue. The solution is given by
\begin{eqnarray}
\varphi_n(y)=N_n e^{2\frac{y}{\ell}}\left[
	J_0(\lambda_n\ell e^{\frac{y}{\ell}})
	-\frac{J_1(\lambda_n\ell)}{N_1(\lambda_n\ell)}
	N_0(\lambda_n\ell e^{\frac{y}{\ell}})\right] \ ,\ \ \ \ 
\end{eqnarray}
where $N_n$ is a normalization constant determined by 
\begin{eqnarray}
\int dy e^{-2{y\over \ell}} \varphi_n (y) \varphi_m (y) = \delta_{nm} \ .
\end{eqnarray}
The junction condition for the negative tension brane gives a
 KK-spectrum as
\begin{eqnarray}
J_1(\lambda_n\ell e^{\frac{d}{\ell}})
	-\frac{J_1(\lambda_n\ell)}{N_1(\lambda_n\ell)}
	N_1(\lambda_n\ell e^{\frac{d}{\ell}})=0 \ .
\end{eqnarray}

Substituting Eq.~(\ref{5:4}) into (\ref{green2}), 
we obtain 
\begin{eqnarray}
&&\hspace{-5mm}
	\delta E_{\mu\nu}(x',y')
	=\int d^4x\sqrt{-g}\sum\limits_n
	\left[
	\varphi_n(d)\varphi_n(y')\Omega^4
 	\hat{\cal L}_{\lambda\rho}{}^{\alpha\beta}
 	\overset{(4)}{G}(x,x')_{\mu\nu}{}^{\lambda\rho}
	\overset{\ominus}{S}_{\alpha\beta}\right.\nonumber\\
&&\left.\hspace{5cm}
	+\varphi_n(0)\varphi_n(y')
 	\hat{\cal L}_{\lambda\rho}{}^{\alpha\beta}
 	\overset{(4)}{G}(x,x')_{\mu\nu}{}^{\lambda\rho}
	\overset{\oplus}{S}_{\alpha\beta}
	\right]  \ .
	\label{green2}
\end{eqnarray}
Using Eq.~(\ref{4d:green}), we can write $\delta E_{\mu\nu}$ as 
\begin{eqnarray}
&&\hspace{-5mm}
	\delta E_{\mu\nu}(x',y')
	=-\sum\limits_n
	\varphi_n(d)\varphi_n(y')\Omega^4
	\overset{\ominus}{S}_{\alpha\beta}
	-\sum\limits_n
	\varphi_n(0)\varphi_n(y')
	\overset{\oplus}{S}_{\alpha\beta}
	\nonumber\\
&&	+\int d^4x\sqrt{-g}\sum\limits_n\lambda_n^2
	\overset{(4)}{G}(x,x')_{\mu\nu}{}^{\lambda\rho}
	\left[\Omega^4\varphi_n(d)\varphi_n(y')
	\overset{\ominus}{S}_{\lambda\rho}
	+\varphi_n(0)\varphi_n(y')
	\overset{\oplus}{S}_{\lambda\rho}
	\right]  \ .
	\label{green2}
\end{eqnarray}
Let us introduce a new  field, which corresponds to a 4-dimensional
massive graviton (KK mode),
\begin{eqnarray}
\hat{\cal L}_{\mu\nu}{}^{\alpha\beta}\psi_{\alpha\beta}^n
	=\mu_n^2\psi_{\mu\nu}^n
\end{eqnarray}
where $\mu_n^2$ is an eigenvalue. Then, the 4-dimensional Green 
function is formally written as
\begin{eqnarray}
\overset{(4)}{G}(x,x';\lambda_m^2)_{\alpha\beta}{}^{\lambda\rho}
       =  \sum\limits_n\frac{\psi^n_{\alpha\beta}(x)\psi_n^{\lambda\rho}(x')}
	{\lambda_m^2-\mu_n^2}
\end{eqnarray}
Note that it is known that this massive graviton causes the Gregory-Laflamme 
instability. Therefore, $\delta E_{\mu\nu}$ also becomes unstable. This implies
that there is a possibility that there exists the scalar wave emition from
the rotating black string due to the Gregory-Laflamme instability.

\section{Conclusion}

We have formulated the black string perturbation theory 
 around the Ricci flat two-brane system. 
 In particular, the master equation for 
 $\delta E_{\mu\nu}$ is derived. 
 At low energy,  the gradient expansion method is utilized to get
 a series solution. This gives the closed system of equations, i.e. 
  the effective Teukolsky equations.
 This can be used for estimating the Kaluza-Klein corrections on 
 the gravitational waves emitted from the perturbed rotating black string.
 Our effective Teukolsky equation is completely separable, hence the
 numerical scheme can be developed in a similar manner as was done 
 in the case of 4-dimensional Teukolsky equation~\cite{Sasaki,Nakamura}.  
 We have also applied our formalism to the Gregory-Laflamme instability.
 We  argued the possibility of scalar wave emition from
 the  black string due to the Gregory-Laflamme instability.

\begin{acknowledgements}
This work was supported in part by  Grant-in-Aid for  Scientific
Research Fund of the Ministry of Education, Science and Culture of Japan 
 No. 155476 (SK) and  No.14540258 (JS) and also
  by a Grant-in-Aid for the 21st Century COE ``Center for
  Diversity and Universality in Physics".  
\end{acknowledgements}


\end{document}